\documentclass[10pt]{iopart}
\usepackage{graphicx}
\begin{document}

\title[Reconstruction of Atomic Ionization Probabilities in Intense Laser Fields]
{Reconstruction of Atomic Ionization Probabilities in Intense Laser Fields}

\author{T R J Goodworth$^1$, W A Bryan$^1$, I D Williams$^2$\\ and W R Newell$^1$}
\address{1) Department of Physics and Astronomy, University College
London, Gower Street, London WC1E 6BT, UK}
\address{2) Department of Physics, Queen's University Belfast,
Belfast BT7 1NN, UK}

\begin{abstract}
Application of a parallel-projection inversion technique to z-scan
spectra of multiply charged xenon and krypton ions, obtained by
non-resonant field ionization of neutral targets, has for the
first time permitted the direct observation of intensity-dependent
ionization probabilities. These ionization efficiency curves have
highlighted the presence of structure in the tunnelling regime,
previously unobserved under full-volume techniques.
\end{abstract}

\section{Introduction}
Multiple atomic ionization within intense ($>$ 10$^{14}$
Wcm$^{-2}$) non-resonant laser fields is an area of considerable
interest [1] involving non-linear phenomena such as
above-threshold ionization, non-sequential ionization [2], and
high harmonic generation [3]. Early investigations collected ion
signal from the full confocal volume, see for example [4]. As
such, detected ion yields represented a convolution of all active
intensity-dependent ionization processes. The use of a full
confocal volume also meant that the ion yield curves as a function
of peak laser intensity followed the characteristic I$^{3/2}$
behaviour within the saturated field ionization regime. The first
attempt to limit this gross spatial integration used a pinhole
aperture to image the centre of a shallow laser focus [5]. Under
this configuration intensity variation was achieved by
conventional half-wave plate energy attenuation techniques.  The
generalised dependence of total ionization has also been studied
by soft focusing the laser beam into an approximately cylindrical
configuration as viewed by the ion detector [6]. More recently,
intensity selective scanning has been successfully implemented
through z-scanning [7, 8]. This technique uses a slit of width
$\Delta$z to limit the detected intensity range along the laser
propagation (z) axis, in this case the intensity variation is
accessed by translating the laser focus along the laser
propagation (z) axis with respect to the masking aperture using
the natural, on-axis (Lorentzian) intensity variation I$_z$ =
I$_0$/(1 + (z/z$_o$)$^2$), where z$_o$ = $\pi
\omega_0$$^2$/$\lambda$ is the Rayleigh range. This technique also
has the advantage that lower intensity processes are accessed with
increased interaction volumes due to the shape of the confocal
volume, see for example [9]. Despite the quantisation of the
intensity variation along the z-axis with this technique, the
radial intensity variation remains unbound. As such, the powerful
z-scan technique is restricted to the determination of threshold
(I$_{th}$) and saturation (I$_{sat}$) intensities by Gaussian
volume fitting [8]. Clearly, if the radial intensity variation
were quantised in a similar way to the z-axial variation, the
z-scan technique would yield complete intensity-dependent
ionization curves for all observed charge states, and also allow a
range of laser intensities to be investigated simultaneously. In
light of this, we present a technique for the quantisation of the
confocal volume to the order ($\Delta$z)$^2$ using a
parallel-projection inversion algorithm in conjunction with the
z-scan technique.
\begin{figure}
\begin{center}
\includegraphics[width=320pt]{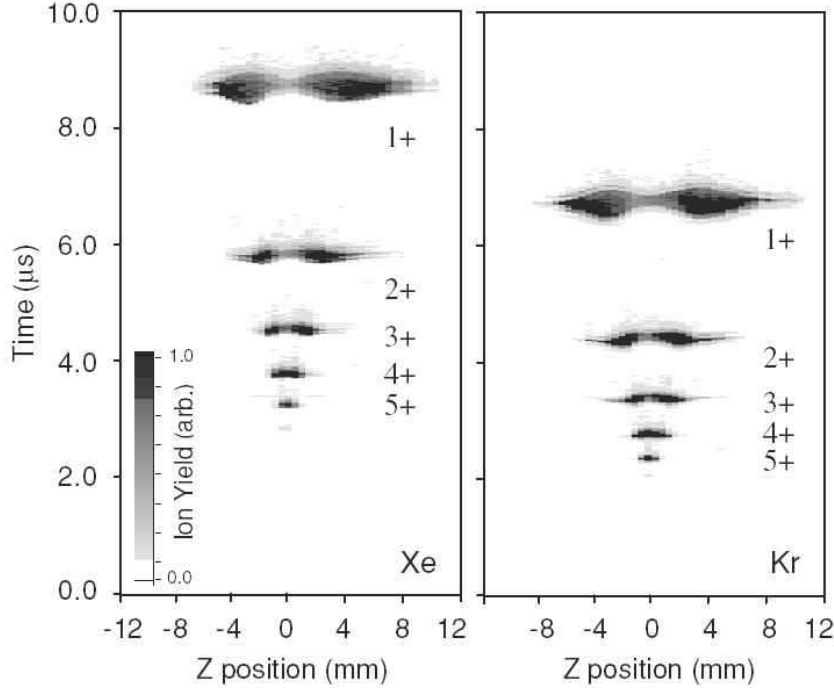}
\caption{Z-scan time-of-flight matrices for Xe and Kr targets.
Centrally located on each figure, the z = 0 axis corresponds to
the position of the peak laser intensity I$_o$. Ion yield profiles
are generated by temporal integration of these plots.
}\label{fig1}
\end{center}
\end{figure}

\section{Inversion Methodology}
Inversion techniques allow determination of an object's internal
properties through the deconvolution of a set of line integrals
intersecting the object, perhaps the most familiar example of
which is computer-assisted tomography (CAT) [10]. Generally, where
the structure of the object is unknown, line integrals from a
number of locations around the object are necessary to obtain the
reconstructed image via an inversion technique such as the Radon
transform [11]. However, if the structure of the object is known,
or has rotational symmetry about an axis normal to the line
integrals, the object may be reconstructed from observations at a
single orbital location. In this instance, the object is the laser
confocal volume, with the line integrals determined by integration
of the rare gas target experimental z-scans which are shown in
figure 1.

In this case, each of the n z-axial measurements ($\eta$$_n$)
represents the total integrated ion yield from all ionization
processes within the discrete intensity range I(z) to I(z+
$\Delta$z) along the z-axis, and an unbound range of intensities
in the radial plane. By constructing a basis set of n Gaussian
isointensity contours I$_s$, each separated by the imaging slit
width $\Delta$z along the z-axis, the entire confocal volume can
be quantised to the order ($\Delta$z)$^2$ by known volume elements
V$_{n,s}$, where n denotes the z-axial position and s refers to
the isointensity contour. These quantised volume elements are
aligned such that the boundary of each contour coincides with the
edge of the imaging slit as shown in figure 2. Importantly, volume
elements common to a particular isointensity shell will possess
the same intensity I$_s$ irrespective of the z-axial position (n).
The functional form of Is is given in equation 1 where r is radial
coordinate orthogonal to z, and $\omega_o$ is the focused laser
beam radius at the position z = 0.
\begin{equation}
I_{s}(z, r)=\frac{I_o}{1 + (z/z_o)^2}\exp \left(\frac{-2r^2}{(1 +
(z/z_o)^2)\omega_o^2}\right)
\end{equation}
\begin{figure}
\begin{center}
\includegraphics[width=280pt]{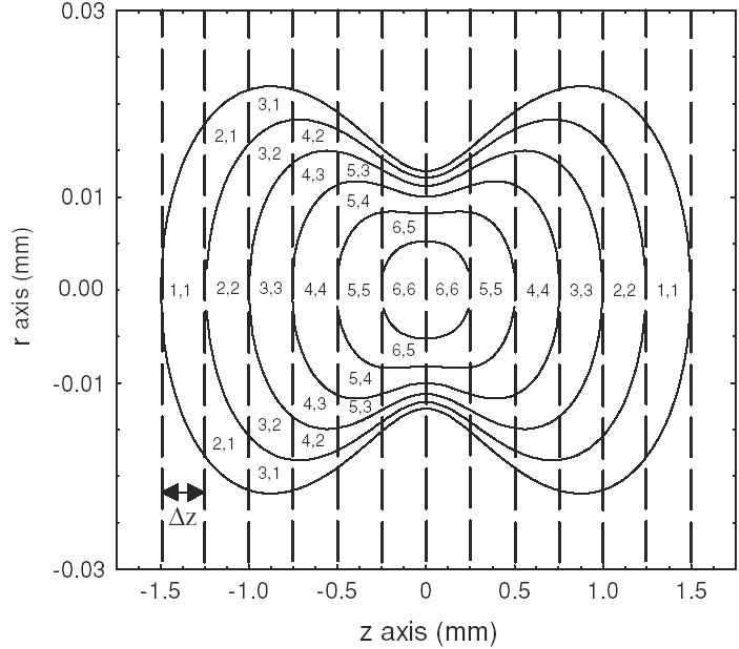}
\caption{Quantisation of the confocal volume by z-scanning. The
isointensity contours are given by equation 1, and the double
indices throughout the figure refer to the volume elements
V$_{n,s}$, where n denotes the slit position along the z-axis, and
s denotes the isointensity contour.}\label{fig2}
\end{center}
\end{figure}
Having fully quantised the confocal volume, the integrated ion
yield profiles can be expressed through the simple matrix product
$\eta_n$ = V$_{n,s}$ $\xi_s$, where $\xi_s$ is the volume
normalised, intensity-dependent ion yield. As the basis sets for n
and s are coincident along the z-axis (see figure 2), the volume
matrix is necessarily well-conditioned, and the system can be
solved immediately for $\xi_s$ by an inversion of the form:
$\xi_s$ = V$^{-1}_{n,s}$ $\eta_n$.

The ion signal profiles used in the inversion algorithm were
obtained from the experimental z-scans depicted in figure 1 by
temporal (vertical axis) integration of the time-of-flight spectra
for each z. This gave an ion signal intensity profile as a
function of z for each observed ion species which in turn acted as
the basis for the inversion process. Diffraction effects present
at very small z values, well above saturation intensities, act to
form additional signal-producing volume over and above that
expected through Gaussian volume calculations [8]. These
additional volume elements were isolated and excluded from the
reconstruction process. In any case the location of this effect,
well above the saturation intensity, places it well out of the
region of interest for the present study.

\section{Experimental Configuration}
The Wiley-McLaren type time-of-flight mass spectrometer used to
collect the z-scan spectra has been described in detail elsewhere
[12], hence only salient points will be discussed here. The
confocal mask which enables z-scanning consists of a 0.5 mm
diameter on-axis aperture situated at the top of a 300 Vcm$^{-1}$
extraction region. For the current optical configuration, this can
be approximated to a semi-infinite slit as the aperture diameter
is much greater than the confocal waist for the full focal
detuning range employed. The target gas is injected directly into
the interaction region at a fill pressure of 10$^{-7}$ mbar on a
background of 10$^{-10}$ mbar, well below the onset of any
observable space-charge effects. At pressures greater than
10$^{-6}$ mbar space charge effects were observed as a widening of
the ion peaks in the TOF spectra and as distortions in the
z-scans. At pressures of 10$^{-7}$ mbar, these effects have
disappeared. The gas is multiply ionized by 65 fs linearly
polarised Ti:Sapphire laser pulses at 10 Hz repetition rate with a
central wavelength of 790 nm, focused using \textit{f}/34 optics
to a peak laser intensity of 7 $\times$ 10$^{15}$ Wcm$^{-2}$. The
laser energy E (= 4 mJ) was measured using a calibrated power
meter and the waist radius $\omega_o$ (= 17 $\mu$m) was calculated
from $\omega_o$ = 2f$\lambda$/$\pi$ where f is the f-number (f =
34) of the optical system. The peak intensity was calculated using
I = E/$\tau\pi\omega_o^2$, where $\tau$ is the laser pulse
duration. The combination of errors associated with meter
calibration and pulse duration measurement give an absolute error
on the intensity of 30\%, although the relative intensities are
accurate to within 10\%. The cylindrical symmetry of the focused
beam is essential preserved in the present optical configuration
since the maximum angle of incidence of the laser beam is less
than 0.5 degrees from the optical axis. This means transmission
through the shallow focusing lens is essentially independent of
the orientation of the polarisation vector with respect to the
plane of incidence [13]. Ions produced during the laser
interaction are detected by a microchannel plate detector
outputting to a fast digital storage oscilloscope, where spectra
are averaged over 500 shots. Shot to shot energy fluctuations were
minimised by a discriminating photodiode. The focal lens was
mounted on a computer controlled precision translation stage,
allowing the focal detuning to be driven with a 0.25 mm step size
with respect to the z-axis, permitting the production of
50-element z-scans. Figure 1 shows the raw z-scan spectra for
xenon and krypton targets. Ions up to Xe$^{5+}$ and Kr$^{5+}$ were
observed, and have been found to be in good agreement with other
experiments employing rare gas targets at these intensities [4,
14].
\begin{figure}
\begin{center}
\includegraphics[width=300pt]{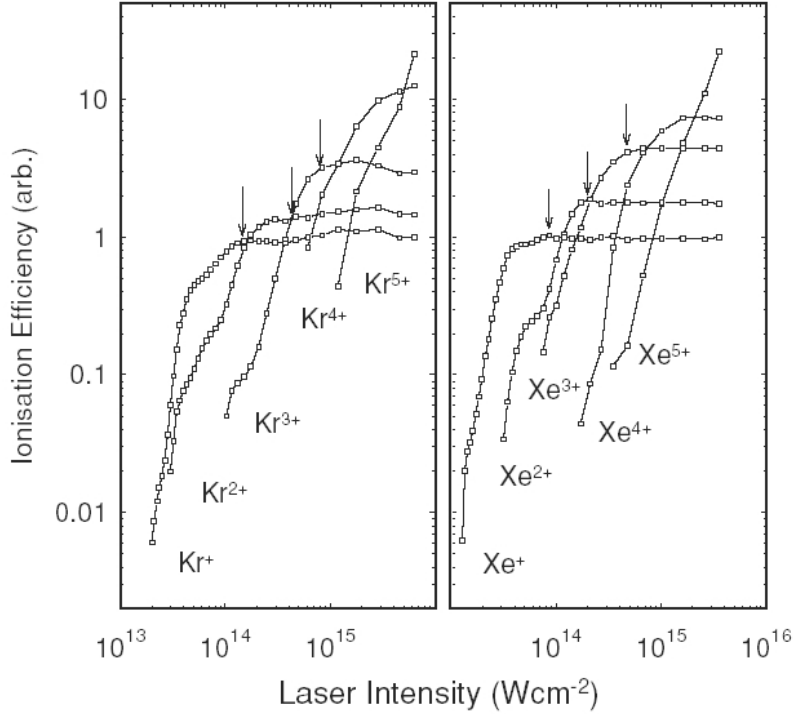}
\caption{Reconstructed intensity-dependent, volume-normalised, Pq
curves as a function of the discrete on-axis local peak intensity
I$_s$ for all observed charge states. The arrows indicate the
saturation intensity expected under an ADK
approximation.}\label{fig3}
\end{center}
\end{figure}
\section{Results and Discussion}
It should be noted that the reconstruction method employed will
yield ionization probability (P) multiplied by the inherent
quantum efficiency (q) of the detector. Following application of
the reconstruction algorithm, the derived Ionization Probability x
Quantum Efficiency (q) is mapped onto an intensity axis as shown
in figure 3. In doing so it is clear that the intensity-dependant
Pq curves for each observed ion species are similar. Each curve is
initiated at a threshold intensity I$_{th}$, followed by a rapid
rise in the Pq characteristic of Stark-shifted multiphoton
ionization. In turn, this is followed by a slow rate of increase,
characteristic of tunnelling ionization until the Pq curve
saturates. Considering the first ionization stage (Xe
$\rightarrow$ Xe$^+$) as an example, it should be noted that the
probability of observing Xe$^+$ would actually decrease at
intensities above saturation as multiple ionization becomes
significant. However the probability of Xe$^+$ being generated
during the pulse is still Pq = 1, and since it provides a more
informative insight into the field ionization dynamics, we have
plotted this probability of production, as opposed to the
probability of observation, in figure 3. The higher ionization
stages display correspondingly higher values of Pq at which they
saturate as a consequence of the varying quantum efficiency of the
detector as a function of charge state. Indeed, measurement of the
Pq values at saturation for multiply charged ions (figure 3)
relative to the singularly charged ion Pq curve provides a direct
experimental measure of the detector quantum efficiency. All Pq
curves in figure 3 can therefore be normalized to a unit
probability (P = 1) at saturation. The Pq curves for the Xe$^{4+}$
and Kr$^{5+}$ are not saturated.

The Coulomb potential plus the laser electric field
($\varepsilon$) gives rise to a total potential of V(r) =
Zr$^{-1}$ - $\varepsilon$.r, where Z is the residual charge on the
ion, and r is the radial distance from the charge centre. A saddle
point in V(r) occurs for $\varepsilon$ = $\varepsilon$$_{sat}$ =
Z$^3$/(4n*)$^2$ [15], where n* (= n - $\delta_l$) is the effective
principal quantum number, and $\delta_l$ denotes the quantum
defect [16]. The calculated $\varepsilon$$_{sat}$ values, which
are both charge and species specific, (shown by arrows in figure
3) are in excellent agreement with the present experimental
saturation intensities. The tunnelling regions in figure 3, in
particular for Kr+ and Xe+, display detail previously unseen in
full confocal volume experiments. Indeed, this technique has, for
the first time, allowed accurate determination of the gradient in
this transitory regime to be made, which perhaps surprisingly
displays an unexpected level of structure, including a
species-specific gradient with Kr$^+$ having a slope three times
that of Xe$^+$. The explanation for this structure requires
further investigation, but we may hypothesize as to the cause.
This is the intensity region where rescattering of the ionised
electron from the singly charged core has been shown to contribute
to non-sequential double ionisation, probably via excitation
followed by field ionisation [17]. It is quite likely that as well
as ionisation, recombination, or indeed field-stimulated
recombination is also occurring in the electron-ion interaction.
This would have the effect of depleting the yield of singly
charged ions, and hence could account for the observed structure.

Using perturbation theory, the low intensity Stark-shifted
multiphoton region can be quantified by I$^N$$\sigma$ where N is
the number of photons and $\sigma$ is the ionization
cross-section. A $\chi^2$ minimisation in this region yielded N =
6(8) for Xe, and N = 7(10) for Kr, less than the expected values
shown in brackets, suggesting that the atomic potential is
strongly influenced by dynamic Stark shifts of 3-5 eV or
alternatively by transient population trapping [18].

The Xe$^{2+}$ and Kr$^{2+}$ Pq curves both display a
characteristic knee structure separating the regions of
non-sequential (NS) and sequential (S) ionization. The transition
from NS to S ionization is completed by laser intensities of 6
$\times$ 10$^{13}$ and 8 $\times$ 10$^{13}$ Wcm$^{-2}$ in
Xe$^{2+}$ and Kr$^{2+}$ respectively. The source of the NS signal
below the knee has previously been attributed to a recollision
mechanism [19]. At 6 $\times$ 10$^{13}$ and 8 $\times$ 10$^{13}$
Wcm$^{-2}$, the corresponding maximum recollision energies are
12.9 and 16.6 eV, leaving a deficit of 8.3 and 7.8 eV against the
ionization potentials of Xe$^+$ and Kr$^+$ respectively,
indicative of excited states undergoing field or electron impact
ionization. Recently, Becker and Faisal [20] have completed
S-matrix calculations in which the primary ionized electron is
dressed by the radiation field [21], which demonstrates the role
of the laser field in electron impact ionization process.
\begin{figure}
\begin{center}
\includegraphics[width=250pt]{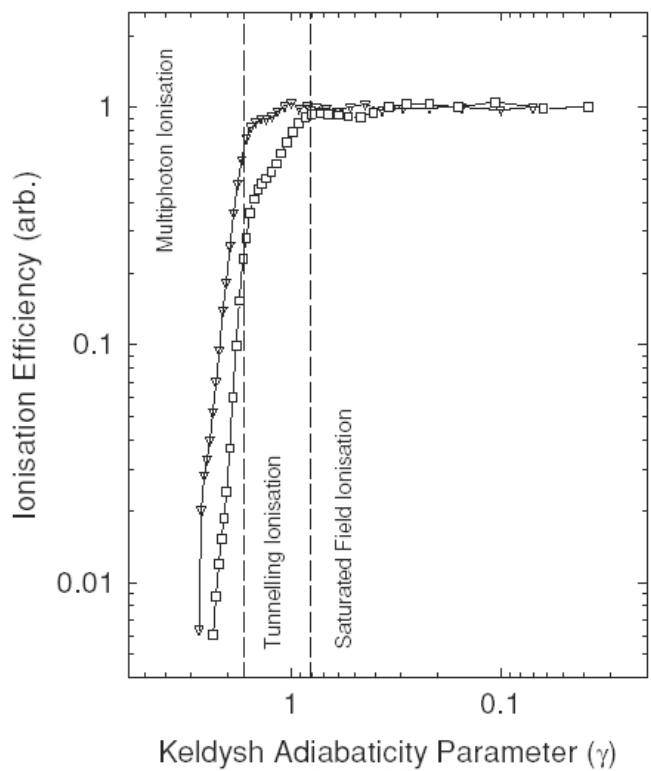}
\caption{Intensity-dependent ionization probabilities for Xe$^+$
($\bigtriangledown$) and Kr$^+$ (  ) as a function of the Keyldsh
parameter. The dashed lines divide the three key regions of atomic
ionization.}\label{fig4}
\end{center}
\end{figure}
A more quantitative interpretation of the intensity-dependent Pq
curves can be obtained by recasting the reconstructed curves in
terms of the Keyldsh adiabaticity parameter $\gamma$ [22], given
in a convenient form by $\gamma$ = (E$_i$/2U$_P$)$^{1/2}$, where
E$_i$ is the ionization potential of the target, $\lambda$ is the
wavelength in microns of the incident radiation field, and U$_P$
is the ponderomotive or quiver energy of the electron in the
field, calculated using U$_P$ = 9.33 $\times$ 10$^{-14}$
$\lambda^2$ I (Wcm$^{-2}$). As shown in figure 4, the curves for
both Xe$^+$ and Kr$^+$ overlap well in $\gamma$-space, with
saturated field ionization occurring for $\gamma$ $<$ 0.8, and the
quantum mechanical tunnelling region straddling $\gamma$ = 1, in
excellent agreement with the theoretically predicted range of 0.5
$<$ $\gamma$ $<$ 1 [15].

\section{Conclusion}
In conclusion, it has been shown that the use of a
parallel-projection inversion algorithm has realised the full
potential of the experimental z-scan technique by, for the first
time, fully quantising the confocal interaction region, allowing
production of volume normalised, intensity-dependent ion yield
curves for a range of charge states in xenon and krypton.

The reported experiment is the first direct observation of
intense-field atomic ionization efficiencies, independent of both
radial intensity blurring and interaction volume effects, which
would usually mask the subtle features detected in figure 3. With
application to molecular targets, this novel technique will permit
an insight into a wide range of intensity-dependent dissociation
and ionization mechanisms with greater clarity.

\ack{We wish to thank the staff of the ASTRA laser facility at the
Rutherford Appleton Laboratory, A J Langley, C J Hooker and E J
Divall for their expert assistance in the collection of
experimental data. This work was supported by the Engineering and
Physical Sciences Research Council. TRJG and WAB gratefully
acknowledge EPSRC awards.}

\section*{References}

\begin{flushleft}
[1] Agostini P et al, Phys. Rev. Lett. 42 (1979) 1127, Yamakawa K
et al, Phys. Rev. Lett. 92 (2004) 123001, Rudenko A et al, Phys.
Rev. Lett. 93 (2005) 253001

[2] Larochelle S, Talebpour A and Chin S L, J. Phys. B: At. Mol.
Opt Phys. 31 (1998) 1201, and references therein.

[3] L'Huillier et al, Atoms in Intense Laser Fields Ed. M Gavrila
(Academic, New York) 1992

[4] Augst S, Meyerhofer D D, Strickland D and  Chin S L, J. Opt.
Soc. Am. B 8 (1991) 858

[5] Jones R R, Phys. Rev. Lett. 74 (1995) 1091

[6] Hankin S M, Villeneuve D M, Corkum P B and Rayner D M, Phys.
Rev. A. 64 (2001) 013405

[7] Hansch P, Walker M A and Van Woerkom LD, Phys. Rev. A. 54
(1996) 2559

[8] El-Zein A et al. Phys. Scripta T92 (2001) 119

[9] Greenwood, J et al. Phys. Rev. Lett. 88 (2002) 233001

[10] Herman G T, Image Reconstruction from Projections, Academic
Press (1980)

[11] Radon J, Math. Phys. 69 (1917) 262

[12] Sanderson J H et al, Phys. Rev. A. 65 (2002) 043403

[13] Ditchburn R W, Light, Blackie (1963)

[14] Talebpour A et al, J. Phys. B: At. Mol. Opt. Phys. 30 (1997)
1721

[15] Ilkov F A, Decker J E and Chin S L, J. Phys B: At. Mol. Opt.
Phys. 25 (1992) 4005

[16] Bethe H A and Salpeter E E, Quantum Mechanics of One and Two
Electron Atoms, Springer (1957)

[17] Feuerstein B et al, Phys. Rev. Lett. 87 (2001) 043003

[18] Talebpour A, Chien C Y and Chin S L, J. Phys. B: At. Mol.
Opt. Phys. 29 (1996) 5725

[19] Corkum P B, Phys. Rev. Lett. 71 (1993) 1994

[20] Becker A and Faisal F H M, J. Phys. B: At. Mol. Opt. Phys. 32
(1999) L335

[21] Van der Hart H W, J. Phys. B: At. Mol. Opt. Phys. 34 (2001)
L147

[22] Keyldsh L V, J. Exp. Theor. Phys. 47 (1964) 1945
\end{flushleft}

\end{document}